\documentclass[showkeys,showpacs,prc,nofootinbib,preprint]{revtex4}
\usepackage{exscale}
\usepackage[dvips]{graphicx}
\usepackage{amssymb,amsbsy}
\newcommand{\bea}{\begin{eqnarray}}
\newcommand{\eea}{\end{eqnarray}}
\newcommand{\bc}{\begin{center}}
\newcommand{\ec}{\end{center}}
\newcommand{\bfi}{\begin{figure}}
\newcommand{\efi}{\end{figure}}
\newcommand{\igr}{\includegraphics}
\newcommand{\sh}{\not\! }

\begin{document}

\title{The spectral function of the $\omega$ meson in nuclear matter from a coupled-channel resonance model}
\thanks{Supported by DFG}
\author{P.~Muehlich}
\email{pascal.muehlich@theo.physik.uni-giessen.de}
\author{V.~Shklyar}
\author{S.~Leupold}
\author{U.~Mosel}
\author{M.~Post}
\affiliation{Institut f\"ur Theoretische Physik, Universit\"at
Giessen, D--35392 Giessen, Germany}

\begin{abstract}
We calculate the spectral function of the $\omega$ meson in nuclear matter at
zero temperature by means of the low-density theorem. The $\omega N$ forward scattering
amplitude is calculated within a unitary coupled-channel effective Lagrangian model
that has been applied successfully to the combined analysis of pion- and photon-induced
 reactions. While the peak of the $\omega$ spectral distribution is
shifted only slightly, we find a considerable broadening of the
$\omega$ meson due to resonance-hole excitations. For $\omega$
mesons at rest with respect to the surrounding nuclear medium, we
find an additional width of about $60~\mathrm{MeV}$ at saturation
density.
\end{abstract}

\maketitle

\section{Introduction}

In the past years substantial theoretical and experimental effort has been directed
to the search for modifications of hadrons embedded in a strongly interacting
environment. These investigations have been driven by the expectation to gather
information about one of the most exciting aspects of quantum chromodynamics (QCD),
namely the restoration of the spontaneously broken chiral symmetry at finite
temperatures and densities. The scalar quark condensate $\langle\bar qq\rangle$,
which plays the role of an order parameter of the symmetry breaking mechanism and
which develops a non-vanishing expectation value in vacuum, is expected to change
from its vacuum value by roughly $30\%$ already at normal nuclear matter density
$\rho_0=0.16~\mathrm{fm}^{-3}$
\cite{Drukarev:1988kd,Cohen:1991nk,Li:1994mq,Brockmann:1996iv}. The prediction
of dropping hadron masses in the nuclear medium driven by the chiral
quark condensate in \cite{Brown:1991kk,Hatsuda:1991ez} has therefore
initiated a wealth of experiments searching for these changes in
various kinds of nuclear reactions. In this respect, the light
vector mesons play a central role as they couple directly to virtual
photons. The latter can decay to dileptons which leave the strongly
interacting system untouched, hence carrying information about the
properties of the decayed vector mesons to the detectors without
experiencing further interactions.

One approach which aims at a connection between hadronic properties
and their in-medium modifications on one hand and non-perturbative
quark and gluon condensates and their in-medium changes on the other
is provided by the QCD sum rule approach. Originally the sum rules
were introduced for the vacuum \cite{shif79} but later on
generalized to in-medium situations \cite{Bochkarev:1986ex}.
Concerning vector mesons it turned out that their in-medium changes
are not directly connected to changes of the two-quark condensate,
but to specific moments of the nucleon structure function and to
changes of the four-quark condensate \cite{Hatsuda:1991ez}. One
might even turn the argument around and state that a measurement of
modified vector meson spectra might give information on the density
dependence of certain four-quark condensates rather than on the
genuine chiral condensate $\langle\bar qq\rangle$
\cite{Zschocke:2002mp,Thomas:2005dc}. Concerning the possible connection of four-
and two-quark condensates see also \cite{Leupold:2005eq}. In
addition, it is important to stress that the sum rule approach does
not directly constrain specific properties of a hadron like e.g.~its
mass. It constrains instead specific integrals over the spectral
(vacuum or in-medium) information. For example, for the $\rho$-meson
one can deduce from sum rules only that in a nuclear medium its
spectral strength is shifted to lower invariant masses. Whether this
shift is realized by a mass shift \cite{Hatsuda:1991ez}, by an
enlarged width \cite{Klingl:1997kf,Leupold:1998dg} or by the
excitation of low-lying resonance-hole pairs \cite{Leupold:2004gh}
cannot be decided from a sum rule analysis alone. Therefore,
hadronic models are needed to describe the specifics of in-medium
changes of vector mesons.

The first experimental observations of a significant reshaping of
the spectral function of the $\rho$ meson have been made by the NA45
\cite{Agakishiev:1995xb,Agakishiev:1997au,Lenkeit:1999xu,Wessels:2002ha}
and HELIOS \cite{Masera:1995ck} collaborations by a measurement of
the dilepton invariant mass spectrum obtained in heavy ion
collisions. The dilepton spectra show a considerable enhancement
over the standard hadronic cocktail calculations consistent with
spectral strength moving downward to smaller invariant masses
\cite{Li:1995qm,Cassing:1995zv,Cassing:1996km,Bratkovskaya:1996qe}.
More recently the NA60 collaboration has obtained dilepton spectra
in heavy-ion collisions with unprecedented mass resolution
\cite{Arnaldi:2006jq}; these results point to a considerable
broadening of the $\rho$ meson, but no mass shift.

The observed signals in heavy-ion collisions necessarily represent
time-integrals over very different stages of the collision (from
initial non-equilibrium states over equilibrium in the QGP and/or
hadronic phase) with significantly changing densities and
temperatures. It has, therefore, been stressed \cite{Mosel:1998rh}
that experiments with elementary probes on ´normal´ nuclei can yield
signals for in-medium changes that are as large as those obtained in
heavy-ion collisions. Such experiments using more elementary
projectiles like photons, protons and pions are of special interest
as they provide an excellent testing ground for theoretical models
based on the assumption of nuclear matter in equilibrium with only
moderate densities ($\leq\rho_0$)
\cite{Falter:2003uy,Alvarez-Ruso:2004ji} (how to go beyond this low
density approximation see \cite{Post:2003hu}).  Following this
proposal, a recent investigation of the dilepton invariant mass
spectrum in $pA$ reactions interpreted in terms of a modification of
the $\rho$ and $\omega$ spectral densities has been made at KEK
\cite{Ozawa:2000iw,Tabaru:2006az,Sakuma:2006xc}. A different
approach has been chosen by the CB/TAPS collaboration by
investigating the $\pi^0\gamma$ channel in $\gamma A$ reactions,
hence focusing exclusively on the properties of the $\omega$ meson
\cite{Trnka:2005ey}, see also \cite{Muhlich:2003tj,Muhlich:2004cm}.
Another promising experiment that is being analyzed by the CLAS
collaboration has taken dielectron data from $\gamma A$ reactions
\cite{Tur:2004}, see \cite{Effenberger:1999ay} for a theoretical
approach.

In the following we concentrate on the $\omega$-meson. (Concerning
theoretical investigations of the $\rho$-meson we refer to
\cite{Rapp:1999ej,Post:2003hu} and references therein.) A lot of
theoretical effort has already been put into the determination of
the isoscalar spectral density in nuclear matter
\cite{Bernard:1988db,Caillon:1995ci,Klingl:1997kf,Friman:1998fb,Klingl:1998zj,Saito:1998wd,Saito:1998ev,Post:2000rf,Lutz:2001mi,Dutt-Mazumder:2002me,Renk:2003hu,Martell:2004gt,Riek:2004kx}.
The outcome of these works covers a rather large area in the mass/width plane, ranging
for the mass from
the free $\omega$ pole mass (782 MeV) down to roughly 640 MeV and for the width
up to about 70 MeV. These approaches differ quite substantially
in their methodical background. Some of them find a shift of
spectral strength to lower invariant masses \cite{Caillon:1995ci,Klingl:1998zj,Saito:1998wd,Saito:1998ev,Dutt-Mazumder:2002me} 
whereas others obtain an upwards shift
\cite{Post:2000rf,Dutt-Mazumder:2000ys,Lutz:2001mi,Zschocke:2002mp,Steinmueller:2006id}. 
For a later comparison to our
result we will pick out two rather different approaches: In the model
of \cite{Lutz:2001mi} the in-medium self energy of the $\omega$ is driven by the
collective excitation of resonance-hole loops. The rather large couplings of
the $D_{13}(1520)$ and $S_{11}(1535)$ nucleon resonances to the $N\omega$ channel
lead to an additional peak in the spectral function, whereas the $\omega$ branch
itself even moves to slightly higher masses due to level repulsion. The model of
\cite{Klingl:1997kf,Klingl:1998zj} is based on an effective Lagrangian which combines
chiral SU(3) dynamics and vector meson dominance. No resonances besides the $\Delta(1232)$
are considered. The authors of \cite{Klingl:1997kf,Klingl:1998zj}
find a rather drastic downward shift of the $\omega$ pole mass with rising
baryon density that can be interpreted as an effect of the renormalization
of the pion cloud generated by the strong interaction between pions and nucleons.

Both approaches \cite{Lutz:2001mi} and
\cite{Klingl:1997kf,Klingl:1998zj} use the fact that to lowest order
in the nuclear density the $\omega$ in-medium self energy is
proportional to the $\omega N$ forward scattering amplitude which is
not directly accessible in experiments.  One solution to this
problem is to use a unitary coupled-channel approach to constrain
the $\omega N$ amplitude. Such calculations have been performed in
\cite{Lutz:2001mi}  where important contributions from nucleon
resonance excitations were found. These resonances were created
dynamically in \cite{Lutz:2001mi} starting from a Lagrangian with
contact interactions. The latter have to be introduced separately
for each partial wave with new input parameters to be fitted to
experimental data. Therefore the authors of \cite{Lutz:2001mi}
restrict themselves to angular momentum $L=0$ between nucleon and
vector meson. This imposes a restriction to the energy region close
to threshold where no contributions from the partial waves with
$L>0$ must be taken into account. On the other hand, this
simplifying approximation that the $\omega$ meson is at rest with
respect to the surrounding nuclear matter is not a valid assumption
for modeling nuclear reactions as the involved momenta of the
produced mesons can acquire quite large values, see e.g.
\cite{Muhlich:2003tj,Muhlich:2002tu}.

In the present work we eliminate this problem by constructing
the  $\omega N$ scattering amplitude from the unitary
coupled-channel $K$-matrix approach developed in \cite{Feuster:1997pq,Feuster:1998cj,%
Penner:2002ma,Penner:2002md,Shklyar:2004dy,Shklyar:2004ba}. This
approach is based on an effective Lagrangian including resonance
fields and aims at a reliable extraction of nucleon resonance
properties from experiments where the nucleon is excited via either
hadronic or electromagnetic probes. The model simultaneously
describes all available data on pion- and photon-induced reactions
on the nucleon for energies $\sqrt{s}\leq 2~\mathrm{GeV}$, including
the final states $\gamma N$, $\pi N$, $2\pi N$, $\eta N$, $\omega
N$, $K\Lambda$ and $K\Sigma$. The same Lagrangian is used for pion-
and photon-induced reactions, allowing for the extraction of a
consistent set of parameters. In contrast to other approaches the
model includes all resonance states with quantum numbers
$J^P=\frac{1}{2}^{\pm}, \frac{3}{2}^{\pm}$ and $\frac{5}{2}^{\pm}$
up to masses of 2 GeV and can thus be used to calculate the self
energy also for finite $\omega$ momenta.

The paper is organized as follows: in Section \ref{lowdens}
we outline the calculation of the $\omega$ in-medium self energy.
Section \ref{kmatrix} gives an overview over the coupled-channel
resonance model and its application to the case at hand.
In Section \ref{sect3} we explore the role of resonance contributions
and coupled-channel effects.
The results are discussed in Section \ref{results}. Finally we summarize
our findings in Section \ref{summary}. In the Appendices some technical
details of the calculations are given.


\section{The model}\label{model}

\subsection{Vector mesons in nuclear matter}\label{lowdens}

The properties of a vector meson at finite nuclear density are characterized by its spectral function $\mathcal{A}_V$ which is the imaginary part of the retarded in-medium vector meson propagator:
\bea\label{spectral}
\mathcal{A}^{T/L}_V(q)=-\frac{1}{\pi}\mathcal{I}m~D_V^{T/L}(q) & = & -\frac{1}{\pi}\mathcal{I}m~ \frac{1}{q^2-m_V^2-\Pi_{\mathrm{vac}}(q)-\Pi^{T/L}_{\mathrm{med}}(q)} \\
& = & -\frac{1}{\pi}\frac{\mathcal{I}m\Pi^{T/L}_{\mathrm{tot}}(q_0,{\bf q})}{\left(q_0^2-{\bf q}^2-m_V^2-\mathcal{R}e\Pi^{T/L}_{\mathrm{tot}}(q_0,{\bf q})\right)^2+\left(\mathcal{I}m\Pi^{T/L}_{\mathrm{tot}}(q_0,{\bf q})\right)^2} \nonumber
\eea
with the vector meson four momentum $q=(q_0,{\bf q})$. In the following we are only concerned with the in-medium self energy $\Pi_{\mathrm{med}}$ and will therefore drop the index $\Pi_{\mathrm{med}}=\Pi$. The vacuum self energy $\Pi_{\mathrm{vac}}$ entering the $\omega$ spectral function will be discussed briefly in Appendix \ref{vacuum}.

The self energies $\Pi^{T/L}(q)$ depend independently on both variables $q_0$ and $|\bf q|$. The indices $^T$ and $^L$ denote the projections on the transverse and longitudinal modes of the vector meson which are obtained by contracting the self energy with the three-transverse and the three-longitudinal projectors:
\bea\label{project}
\Pi^T & = & \frac{1}{2}P_T^{\mu\nu}\Pi_{\mu\nu} \,, \nonumber\\
\Pi^L & = & P_L^{\mu\nu}\Pi_{\mu\nu}.
\eea
For a definition of $P_T^{\mu\nu}$ and $P_L^{\mu\nu}$ see e.g. \cite{Post:2003hu}.

\bfi[hbt]\bc
\igr[scale=1.]{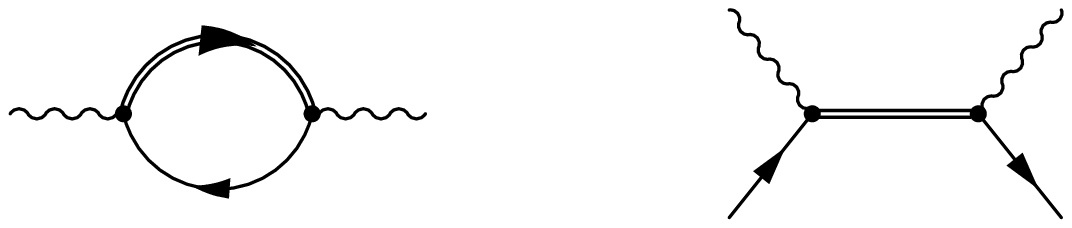}\\
\hspace*{.4cm}(a)\hspace{6.4cm}(b)
\caption{Lowest order contribution to the vector meson self energy in nuclear matter (a)
and vector meson nucleon scattering via excitation and decay of an $s$-channel
resonance (b).}
\label{selfen}
\ec\efi

In the present work we use the low-density theorem \cite{Dover:1971} to determine
the in-medium self energy of the $\omega$-meson. We start by illustrating this theorem
by a specific (and important) type of process:
From the coupling of the vector meson to baryon resonances, the contribution to the
in-medium self energy $\Pi_{\mu\nu}$ is given to lowest order by the diagram depicted
in Fig.~\ref{selfen}(a). Applying standard Feynman rules, considering for illustration
a resonance $R$ with spin $\frac{1}{2}$ and neglecting the $u$-channel contribution,
one arrives at the following expression \cite{Peskin:1995ev}:
\bea\label{pimunu1}
i\Pi_{\mu\nu}(q)=(-ig)^2(-1)\int\frac{d^4p}{(2\pi)^4}\mathrm{Tr}\left\{\Gamma_{\mu}G_N(p)\Gamma_{\nu}G_R(k)\right\},
\eea
where we have assumed a not further specified coupling at the $RNV$ vertex of the form
\bea
\mathcal{L}_{RNV}\sim g\bar u_R\Gamma_{\mu}u_NV^{\mu}.
\eea
The momentum of the resonance is $k=p+q$. $G_N$ and $G_R$ denote the (in-medium) propagators of the nucleon and the spin-$\frac{1}{2}$ resonance. By inserting the relativistic in-medium nucleon propagator
into Eq.~(\ref{pimunu1}) and keeping only the relevant part corresponding to the propagation of nucleon holes, the above expression simplifies to
\bea\label{pimunu2}
\Pi_{\mu\nu}(q_0,{\bf q})=g^2\int\frac{d^3p}{(2\pi)^3}\frac{1}{2E_N({\bf p})}\Theta(p_F-|{\bf p}|)\mathrm{Tr}\left\{\Gamma_{\mu}(\sh p+m_N)\Gamma_{\nu}G_R(k)\right\},
\eea
where we have introduced the nucleon Fermi momentum in the local
density approximation
\bea
p_F=\left(\frac{3}{2}\pi^2\rho\right)^{\frac{1}{3}}~;
\eea
$\rho$ denotes the nucleon density. The trace $\mathrm{Tr}\{\dots\}$
in Eq.~(\ref{pimunu2}) is proportional to the forward scattering
tensor $T_{\mu\nu}$ for the process $V(q)N(p)\rightarrow
R(k=p+q)\rightarrow V(q)N(p)$, depicted in Fig.~\ref{selfen}(b). In
the limit of low densities, the ${\bf p}$-dependence of the
scattering tensor can be neglected, allowing to carry out the
integral explicitly. Hence we arrive at the simple expression:
\bea\label{trho}
\Pi_{\mu\nu}=\rho T_{\mu\nu},
\label{self_energ}
\eea
which is known as the low density theorem \cite{Dover:1971}.
Eq.~(\ref{trho}) is a very general expression which is not restricted to the formation
of resonances. It holds basically for all processes involving one nucleon line at a
given time. To get the complete self energy one needs the complete $\omega$-nucleon
forward scattering amplitude in (\ref{trho}).
In the following section we show how the not directly measurable $\omega N$ forward scattering tensor can be obtained by solving the coupled-channel scattering problem.


\subsection{The $\omega$-nucleon scattering amplitude}\label{kmatrix}

An extensive description of the coupled-channel approach has been given in
\cite{Penner:2002ma,Penner:2002md,Penner:PhD,Shklyar:2004dy,Shklyar:2004ba} and references therein.
Here we briefly outline some of the main features of the model focusing mainly on the
$\omega N$ scattering tensor.

The $2\rightarrow 2$
scattering amplitude is obtained by summing the two-body interaction potential
to all orders, while the physical constraints as relativistic invariance,
unitarity and gauge invariance are preserved. This corresponds to a solution
of the Bethe-Salpeter equation that is shown graphically in Fig.~\ref{bethe}.
Formally, i.e.~dropping the arguments and the integration/summation over
the intermediate states, the Bethe-Salpeter equation can be written as
\bea\label{bse}
M=V+VG_{\mathrm{BS}}M
\eea
with the two-body interaction potential $V$ and the Bethe-Salpeter
propagator
$G_{\mathrm{BS}}$ that is the product of the intermediate state nucleon
and meson propagators. $M$ is the full two-body scattering amplitude
containing also rescattering effects. To solve this equation the
so-called $K$-matrix approximation is applied. Here the real part
of the propagator $G_{\mathrm{BS}}$ is neglected, which corresponds
to putting all intermediate particles on their mass shell. The
asymptotic particle states considered in our approach are
$\gamma N,~\pi N,~2\pi N,~\eta N,~\omega N,~K\Lambda$ and $K\Sigma$.

\bfi[hbt]\bc
\igr[scale=.8]{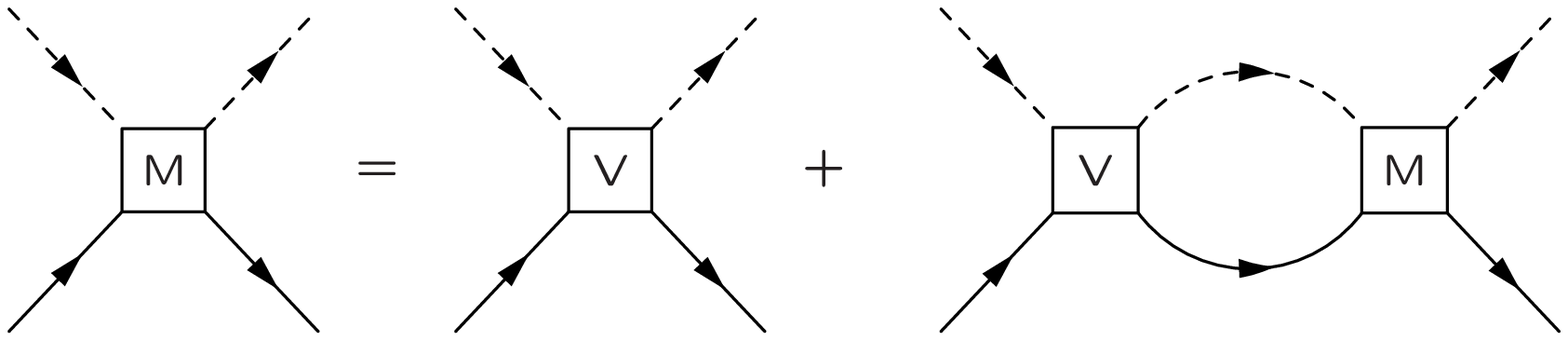}
\caption{Bethe-Salpeter equation for the two-particle scattering amplitude.}
\label{bethe}
\ec\efi

The interaction potential $V$ entering the Bethe-Salpeter equation is
built up as a sum of the $s$-, $u$- and $t$-channel contributions
corresponding to the tree-level diagrams shown in Fig.~\ref{tree}.
The internal lines in the diagrams (a) and (b) represent either a
nucleon or a baryon resonance. In the $t$-channel exchange diagram (c)
the contributions from scalar, pseudoscalar and vector mesons are taken
into account, see e.g.~\cite{Shklyar:2004ba}. Thus, resonance and
background contributions are generated consistently from the same interaction
Lagrangians. The Lagrangians used to construct the $K$-matrix kernel are
given in the literature \cite{Penner:2002ma,Penner:2002md,Shklyar:2004dy,Shklyar:2004ba}.
For completeness, the  resonance $RN\omega$ couplings are summarized
in Appendix~\ref{lagrangians}.
Applying a partial wave decomposition of the scattering amplitudes,
Eq.~(\ref{bse}) can be rewritten in the $K$-matrix approximation in the form
\bea\label{bse2}
T_{ij}^{J^{\pm},I}=K_{ij}^{J^{\pm},I}+i\sum\limits_kT_{ik}^{J^{\pm},I}K_{kj}^{J^{\pm},I},
\label{bethe1}
\eea
where $T_{ij}^{J^P,I}$ is a scattering amplitude for the total spin $J$,
parity $P$ and isospin $I$. The indices $i,j,k$ denote the various
final states $i,j,k=\pi N$, $2\pi N$, $\omega N$, etc.

In such a treatment of the scattering problem the transition amplitudes
$T_{\omega N \to \omega N}^{J^{\pm}}$ are the result of solving the
coupled-channel equation (\ref{bethe1}) where resonance contributions
and rescattering effects are included in a selfconsistent way.
In a previous calculation \cite{Shklyar:2004ba} the updated solution
to the $\pi N \to \gamma N$, $\pi N$, $2\pi N$, $\eta N$, $\omega N$, $K\Lambda$, $K\Sigma$ and
$\gamma  N \to \gamma N$, $\pi N$, $\eta N$, $\omega N$, $K\Lambda$, $K\Sigma$
reactions in the energy region
$\sqrt{s}\leq 2\,$GeV has been obtained. The a priori unknown resonance coupling constants
have been obtained from the fit to  all available experimental reaction
data in the energy region under discussion.
As a  result of these calculations the elastic $\omega N$ scattering
amplitudes of interest have been extracted. Here, we use these amplitudes
as an input for the calculation of the $\omega$ spectral
function at finite nuclear density by means of the low-density theorem.

\bfi[hbt]\bc
\igr[scale=.8]{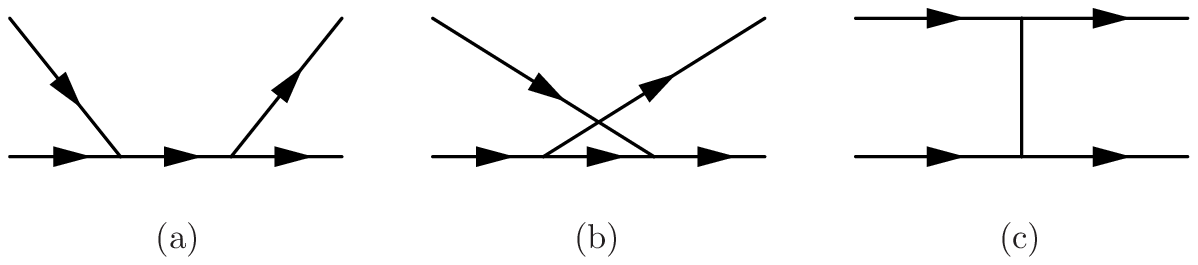}
\caption{$s$-, $u$- and $t$-channel contributions to the interaction kernel.}
\label{tree}
\ec\efi

For the case at hand we have extended our coupled-channel model  in order to
allow for arbitrary  four momenta of the $\omega$ meson as required
by Eq.~(\ref{spectral}). Thus, we are interested in the forward scattering amplitude
as a function of the two independent variables $|\bf q|$ and $q_0$ or,
alternatively, $q^2$. The extension is
achieved by introducing an additional final state into the Bethe Salpeter
equation (\ref{bethe1})
that  we call $N\omega^*$. Such an additional 'effective' $\omega^*$ meson is
characterized by  completely identical properties as the physical
$\omega$ meson apart from its mass $\sqrt{q^2}$ that can take arbitrary values.
To calculate the amplitude $T_{\omega^*N}$ as a function of
$q_{\omega^*}^2=m_{\omega^*}^2\not=m_{\omega}^2$ we follow the procedure used
in \cite{Penner:2002md} to describe photon-induced reactions on the nucleon.
With the amplitudes from Eq.~(\ref{bse2}) we obtain
\bea
T_{\omega^*N,i}^{J^{\pm},I}&=&K_{\omega^*N,i}^{J^{\pm},I}
+i\sum\limits_{j\not=\omega^*N}K_{\omega^*N,j}^{J^{\pm},I}T_{j,i}^{J^{\pm},I},\\
T_{\omega^*N,\omega^*N}^{J^{\pm},I}&=&K_{\omega^*N,\omega^*N}^{J^{\pm},I}
+i\sum\limits_{j\not=\omega^*N}K_{\omega^*N,j}^{J^{\pm},I}T_{j,\omega^*N}^{J^{\pm},I},\label{bse4}
\eea
where the amplitudes $T_{j,i}^{J^{\pm},I}$  are solutions of the coupled-channel
problem taken from \cite{Shklyar:2004ba}. The matrices $K_{\omega^*N,i}^{J^{\pm},I}$
contain the interaction potential for the transitions
$\omega^* N\to \omega^* N$, $\omega N$, $\pi N$, etc., and are chosen to be
the same as $K_{\omega^N,i}^{J^{\pm},I}$ but with $m_{\omega^*}\not  =  m_\omega$.
It is easy to see that for $q_{\omega^*}^2=m_{\omega^*}^2=m_{\omega}^2$
the amplitudes $T_{\omega^*N}$ from Eq.~(\ref{bse4}) and $T_{\omega N}$
in Eq.~(\ref{bse2}) become equal. Note, that introducing the $\omega^*N$
final state does not destroy the unitarity of $T_{\omega N}$ since the
$\omega^* N$ channel does not appear in the intermediate state in Eq.~(\ref{bse4}).
In this way we obtain the vacuum scattering amplitude entering the low
density theorem: The in-medium $\omega$ meson (the outer legs) can take
arbitrary four momenta $(q_0,{\bf q})$, whereas the internal lines
maintain the vacuum properties of the $\omega$ and all other mesons,
i.~e. the four momentum of the internal $\omega$ is constrained by the
on-shell condition $q_0=E_{\omega}({\bf q})=\sqrt{m_{\omega}^2+{\bf q}^2}$.
This corresponds to a first order expression in the nuclear density,
taking into account only interactions with one nucleon at a time.


\section{The role of resonance contributions}\label{sect3}

The $\omega N$ scattering amplitude derived in \cite{Shklyar:2004ba}
and used in the present calculations is a coherent sum of resonance
and background contributions including multi-rescattering effects in
a number of intermediate channels: $\pi N$, $2\pi N$, etc. In
contrast, the work of \cite{Klingl:1997kf,Klingl:1998zj} uses only
Born and mesonic box diagrams with the $\Delta$ being the only
resonance excitation considered. In this Section we, therefore,
illustrate the importance of a single resonance excitation at $\sim
1.65$ GeV and coupled-channel effects. To this end we construct a
simplified model for  $\omega N$ scattering where the transition
amplitude under discussion is a sum of the infinite series of
diagrams shown in Fig.~\ref{toy_model}. It corresponds to solving
the coupled-channel  Bethe-Salpeter equation in the $K$-matrix approximation with
the $\omega N$ and  $\pi N$ channels including also the transitions $\pi N\to \omega N$.
As a showcase we take into account the excitation of the
$S_{11}(1650)$ resonance in the intermediate $\pi N$ channel. The
choice of this resonance is motivated by its mass which is close to
the $\omega N$ threshold. Note that here this resonance contributes to $\omega N$ scattering
only indirectly via the rescattering in the intermediate $\pi N$
channel.

\bfi\bc
\igr[scale=0.8]{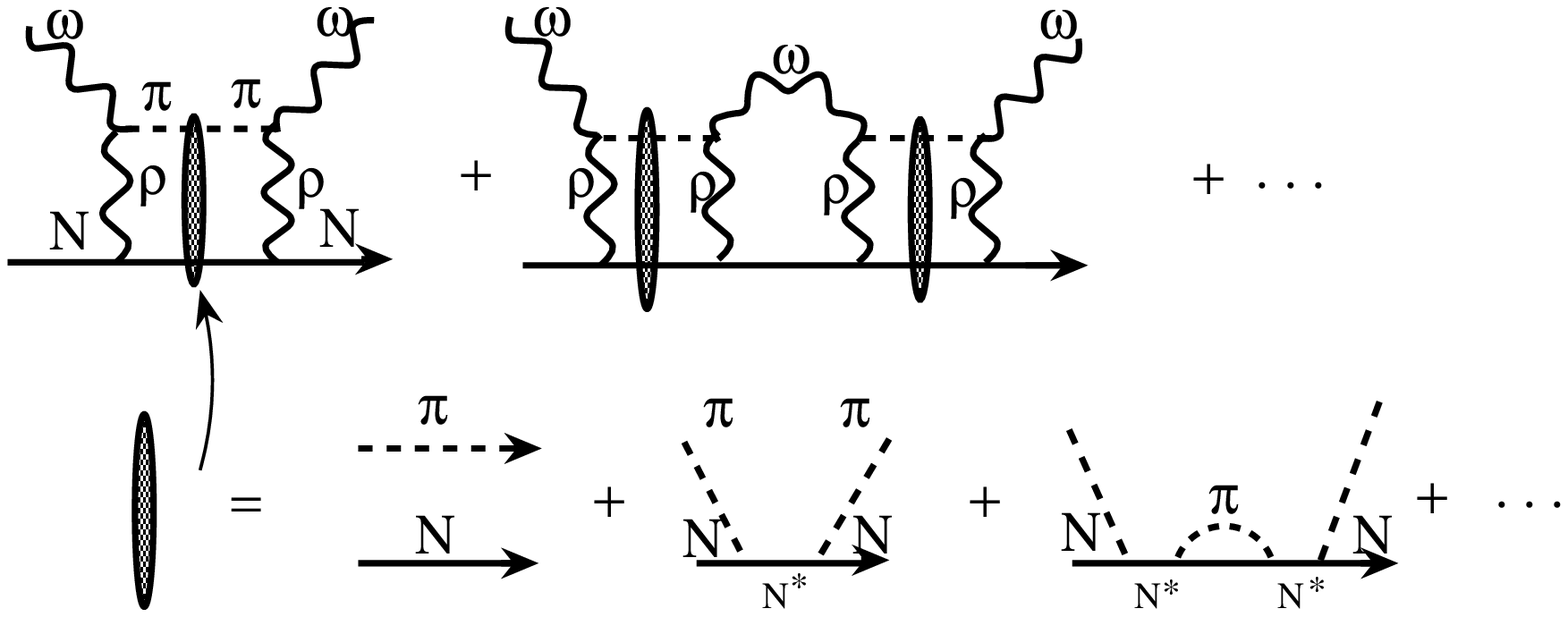}
\caption{Simplified model for $\omega N$ scattering to explore the effect of resonance
excitations. $T_{\omega N}$ is a sum of an infinite series of diagrams (upper panel) with
the $S_{11}(1650)$ resonance excitation in the intermediate $\pi N$ channel. }
\label{toy_model}
\ec\efi

To explore the role of the $S_{11}(1650)$ resonance excitation we
calculate the total cross sections $\sigma_{\omega N}$ using
$\Gamma^{\pi N}_{N(1650)}=0$ and $\sigma_{\omega N}^{N(1650)}$ using
$\Gamma^{\pi N}_{N(1650)}=95$ MeV, respectively. Thus, in the first
case the resonance contributions are absent. The non-resonance
couplings are chosen in accordance with \cite{Shklyar:2004ba}. Form
factors are neglected for the sake of simplicity. To minimize the
dependence on the choice of the coupling constants at the
non-resonance vertices we calculate the ratio of the total cross
sections $\sigma_{\omega N}/\sigma_{\omega N}^{N(1650)}$ that is
shown in Fig.~\ref{ratio} as a function of the c.m. energy. One can
see a dramatic change
\bfi\bc
\igr[scale=1.]{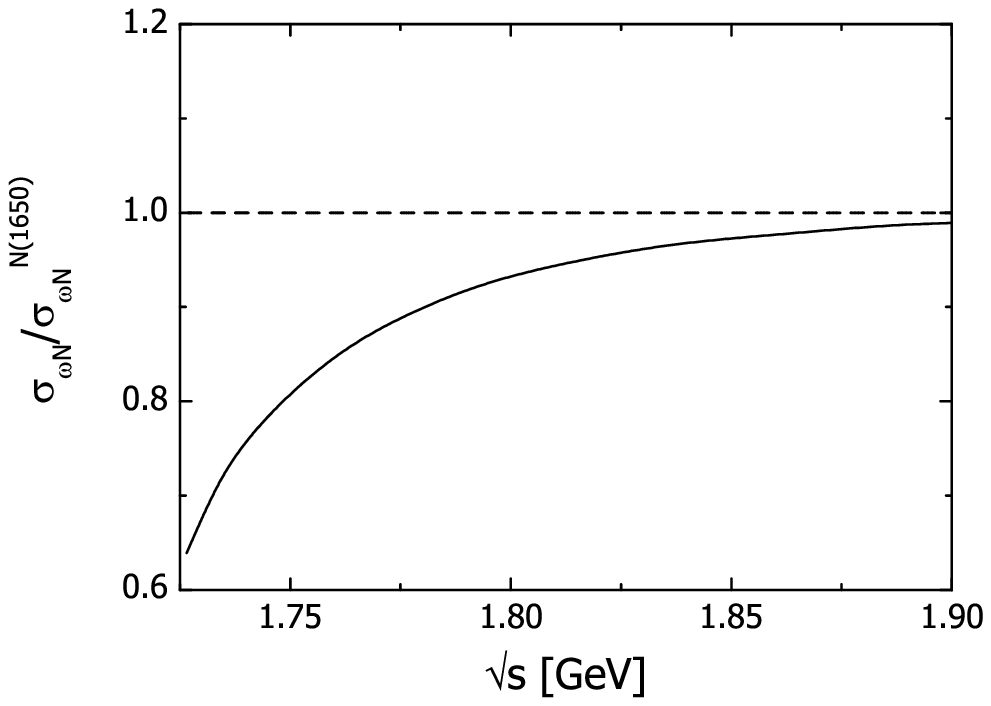} \caption{Ratio of total cross
section calculated within the simplified model displayed in
Fig.~\ref{toy_model} without resonance contributions
($\sigma_{\omega N}$) to that including the $S_{11}(1650)$ state
($\sigma_{\omega N}^{N(1650)}$). } \label{ratio} \ec\efi
of the  $\omega N$ scattering once the contribution from  the $S_{11}(1650)$
resonance is included. As was expected the main difference
between the two calculations is found in the energy region
close to the resonance pole.
At higher energies the contribution from the $S_{11}(1650)$ vanishes and both
results coincide. Note, that no direct resonance couplings
to the $\omega N$ channel are allowed in these calculations. We conclude that
for a realistic description of the $\omega N$ scattering amplitude
coupled-channel effects must be taken into account.
The contributions from nucleon resonances cannot be neglected even
for vanishing resonance couplings to the $\omega N$ final state.


\section{Results}\label{results}

According to the low-density theorem,  the $\omega$ spectral
function is entirely determined by the forward scattering amplitude
$T_{\omega N}$. At zero momentum the latter reduces to the
scattering length. Hence, the  $\omega N$ scattering length
$a_{\omega N}$ defines the $\omega $ meson self energy at the
physical mass. In \cite{Shklyar:2004ba} the scattering lengths and
effective radius have been extracted from the $\omega N$ scattering
amplitude.  Here, we follow \cite{Klingl:1998zj} and define
$a_{\omega N}$ in a slightly different way which is useful for the
present calculations (see Appendix \ref{scattering}). For the
$\omega N$ scattering length we obtain the values
\bea
a_{\omega N}&=&a_{\omega N}^{\frac{1}{2}-}+a_{\omega N}^{\frac{3}{2}-}
= (-0.17+i 0.31)~\mathrm{fm},\nonumber\\
a_{\omega N}^{\frac{1}{2}-}&=&(-0.27 + i0.16)~\mathrm{fm}  ,\nonumber\\
a_{\omega N}^{\frac{3}{2}-}&=&(+0.11 + i0.15)~\mathrm{fm},
\eea
where $a_{\omega N}^{\frac{1}{2}-}$ and $a_{\omega
N}^{\frac{3}{2}-}$ are the contributions  from the spin $\frac12$
($S_{11}$) and spin $\frac32$ ($D_{13}$) sector, respectively. While
we find an attraction in the $D_{13}$ wave the contribution from
$S_{11}$ dominates the real part leading to the slight overall
repulsion in the $\omega N$ system. This result has to be compared
with $a_{\omega N}= (1.6+ i0.3)$ fm obtained by Klingl et el. \cite{Klingl:1998zj}
and $a_{\omega N}= (-0.44+i0.2)$ fm obtained by Lutz et al.
\cite{Lutz:2001mi}. While the imaginary parts in all three
calculations are similar, there is a spread of values in the real
part; we will comment later on this variation.

In Fig.~\ref{spectral_dens}  the $\omega$ spectral function at finite
\bfi\bc \igr[scale=1.]{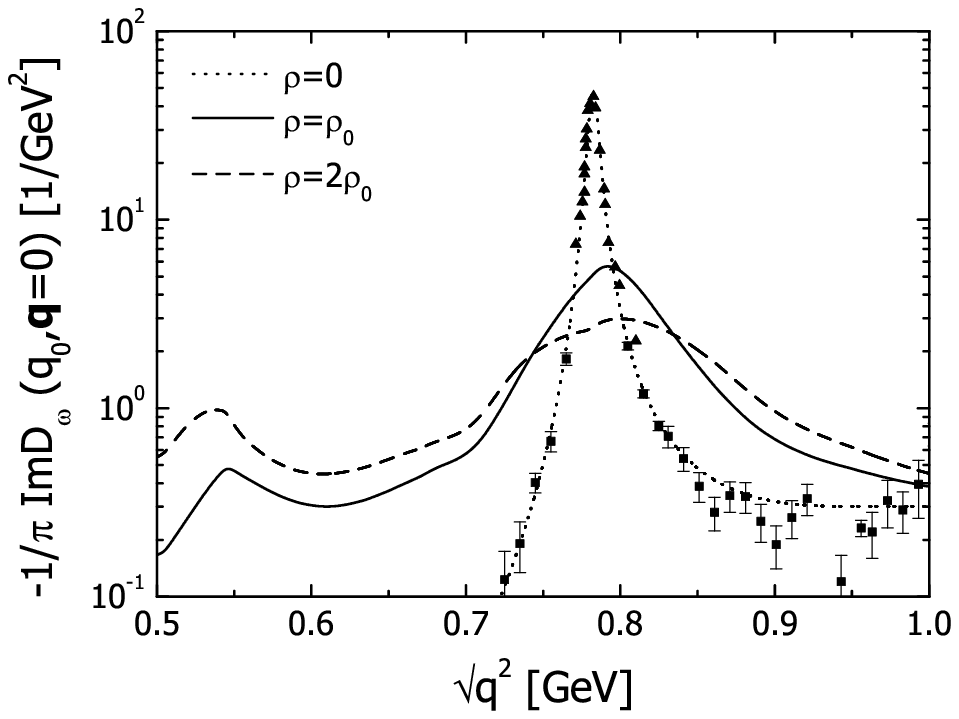} \caption{The $\omega$
spectral function for an $\omega$ meson at rest,
i.e.~$q_0=\sqrt{q^2}$. The appropriately normalized data points
correspond to the reaction $e^+e^-\rightarrow\omega\rightarrow 3\pi$
in vacuum, taken from Ref.~\cite{Barkov:1987ca,Dolinsky:1991vq}.
Shown are results for densities $\rho=0$,
$\rho=\rho_0=0.16~\mathrm{fm}^{-3}$ (solid) and $\rho=2\rho_0$
(dashed).} \label{spectral_dens} \ec\efi
nuclear densities $\rho=0,~\rho=\rho_0=0.16~\mathrm{fm}^{-3}$ and
$\rho=2\rho_0$ is shown for an $\omega$ meson that is at rest with
respect to the surrounding nuclear matter. The appropriately
normalized data points correspond to the process
$e^+e^-\rightarrow\gamma^*\rightarrow\omega\rightarrow\pi^+\pi^-\pi^0$
that directly resembles the $\omega$ vacuum spectral function. Most
noticeable the $\omega$ meson survives as a quasi particle at
nuclear saturation density which is in agreement with all competing
approaches known by the authors. The main effect of the in-medium
self energy is a considerable broadening of the peak that amounts to
roughly 60 MeV at $\rho=\rho_0$. This value is in line with a recent
attenuation analysis \cite{Muhlich:2006ps}. The peak position is
shifted upwards only slightly by about 10 MeV. Due to the collective
excitation of resonance hole loops the spectral function shows a
second peak at low values of the $\omega$ invariant mass
$\sqrt{q^2}$.

The $\omega$ in-medium self energy including the excitation of resonance-hole
pairs exhibits a remarkably rich structure, see Fig.~\ref{self}, where we show
\bfi\bc
\igr[scale=1.4]{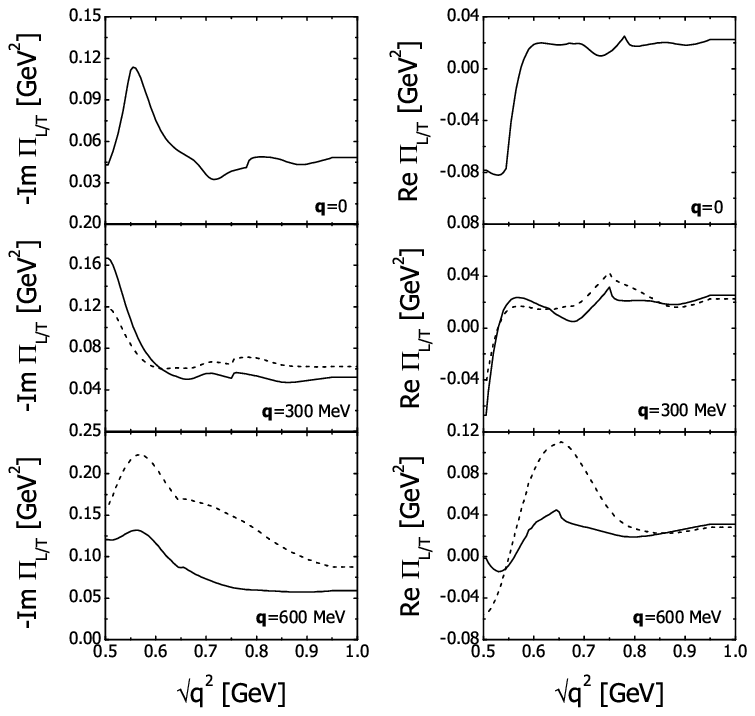}
\caption{Real (right panel) and imaginary (left panel) part of the $\omega$ self
energy in nuclear matter at saturation density $\rho_0=0.16~\mathrm{fm}^{-3}$.
Shown are the longitudinal (solid) and transverse (dashed) modes for $\omega$
three-momenta $|{\bf q}|=0$, $|{\bf q}|=300~\mathrm{MeV}$, $|{\bf q}|=600~\mathrm{MeV}$
with respect to nuclear matter at rest.}
\label{self}
\ec\efi
the real and imaginary part of the $\omega$ self energy of both the longitudinal
and transverse mode for three-momenta $|{\bf q}|=0~\mathrm{MeV},~300~\mathrm{MeV}$
and 600 MeV at normal nuclear matter density. In the limit of very small
resonance widths, each resonance-hole pair generates an additional branch in
the spectral distribution which leads to a multi peak structure. As the widths
of most of the involved resonances for the case at hand are large, see
Table~\ref{table01}, almost no individual peaks can be distinguished and
the resonance excitations add up to a background like structure.
\begin{table}[hbt!]
\bc
\vspace*{.3cm}
\begin{tabular}{l|c|l|l|l|l|l|l|l|l|l}
\hline\hline
$L_{2I,2S}$ & status & mass & $\Gamma_{\mathrm{tot}}$ & $R_{\pi N}$ & $R_{2\pi N}$ &
$R_{\eta N}$ & $R_{\omega N}$ & $g_{RN\omega}^1$ & $g_{RN\omega}^2$ & $g_{RN\omega}^3$ \\
\hline\hline
$S_{11}(1535)$ & **** & 1526 & 136 & 34.4 & 9.5 & 56.1 & -- & 3.79 & 6.50 & -- \\
$S_{11}(1650)$ & **** & 1664 & 131 & 72.4 & 23.1 & 1.4 & -- & $-1.13$ & $-3.27$ & -- \\
\hline
$P_{11}(1440)$ & **** & 1517 & 608 & 56.0 & 44.0 & 2.82 & -- & 1.53 & $-4.35$ & -- \\
$P_{11}(1710)$ & *** & 1723 & 408 & 1.7 & 49.8 & 43.0 & 0.2 & $-1.05$ & 10.5 & --\\
\hline\hline
$P_{13}(1720)$ & **** & 1700 & 152 & 17.1 & 78.7 & 0.2 & -- & $-6.82$ & $-5.84$ & $-8.63$ \\
$P_{13}(1900)$ & ** & 1998 & 404 & 22.2 & 59.4 & 2.5 & 14.9 & 5.8 & 14.8 & $-9.9$ \\
\hline
$D_{13}(1520)$ & **** & 1505 & 100 & 56.6 & 43.4 & 0.012 & -- & 3.35 & 4.80 & $-9.99$ \\
$D_{13}(1950)^a$ & ** & 1934 & 859 & 10.5 & 68.7 & 0.5 & 20.1 & $-10.5$ & $-0.6$ & 17.4 \\
\hline\hline
$D_{15}(1675)$ & **** & 1666 & 148 & 41.1 & 58.5 & 0.3 & -- & 109 & $-99.00$ & 83.5 \\
\hline
$F_{15}(1680)$ & **** & 1676 & 115 & 68.3 & 31.6 & 0.0 & -- & 12.40 & $-35.99$ & $-78.28$ \\
$F_{15}(2000)$ & ** & 1946 & 198 & 9.9 & 87.2 & 2.0 & 0.4 & $-19.6$ & 19.3 & 23.14 \\
\hline\hline
\end{tabular}
\vspace*{.3cm}
\caption{Properties of the $J^P=\frac{1}{2}^{\pm},\frac{3}{2}^{\pm}$ and
$\frac{5}{2}^{\pm}$ resonances that couple to the $N\omega$ channel.
Masses and widths are given in MeV and the on-shell decay ratios $R$
are given in percent. The current status is quoted as in Ref.~\cite{Eidelman:2004wy}.
In addition, also the $RN\omega$ coupling constants entering the Lagrangians
(Appendix \ref{lagrangians}) are given. $^a$: in Ref.~\cite{Eidelman:2004wy}
listed as $D_{13}(2080)$.}
\label{table01}
\ec
\end{table}

However, for the $\omega$ meson at rest one additional peak in both
the imaginary part of the self energy and the spectral function can
be identified at $\sim 0.55$ GeV (see Figs.~\ref{self} and
\ref{spectral_dens}). This branch of the $\omega$ spectral function
is due to the excitation of the $S_{11}(1535)$ resonance. This
resonance couples in relative $s$-wave to the $N\omega$ channel and
dominates the spectrum at low $\omega$ momenta and low $q^2$. Note,
that the authors of \cite{Lutz:2001mi} come to the same conclusion
on the role of the $S_{11}(1535)$ state. However, contrary to
\cite{Lutz:2001mi} we see no prominent effect from  $D_{13}(1520)$
because of the smaller coupling of this resonance to the $\omega N$
final state. The invariant mass $\sqrt{q^2}$ of the  $S_{11}(1535)$
resonance-hole branch moves to smaller values as the three-momentum
increases and can approximately be determined by the kinematical
relation
\bea
(q+p)^2=q^2+m_N^2+2m_N\sqrt{q^2+{\bf q}^2}=m_R^2.
\eea
Therefore, in Fig.~\ref{self} the resonance-hole peak 
visible at a mass of $\sqrt{q^2}\approx550$ MeV
for zero momentum moves down to $\sqrt{q^2}\approx500$ MeV for a momentum of
300 MeV.

Another interesting structure is visible in Fig.~\ref{self} at masses of 782 MeV
($|{\bf q}|=0$ MeV), 750 MeV ($|{\bf q}|=300$ MeV) and 649 MeV
($|{\bf q}|=600$ MeV). This cusp structure is due to the opening of
the elastic channel $\omega^*(\sqrt{q^2})N\rightarrow\omega(782)N$,
i.e.~the scattering of the off-shell $\omega^*$ into the on-shell
$\omega$ becomes energetically possible. The position of this
threshold is determined by the equation
\bea
q^2=m_N^2+(m_N+m_{\omega})^2-2m_N\sqrt{(m_N+m_{\omega})^2+{\bf q}^2}.
\eea
We should note that this threshold effect is an artifact of the applied
low-density approximation. If the self energy was obtained in an
iterative scheme, i.~e. taking into account higher order density effects,
this cusp structure would be smeared out.

In Fig.~\ref{spectral_plab} we show the longitudinal and transverse mode
\bfi\bc
\igr[scale=1.]{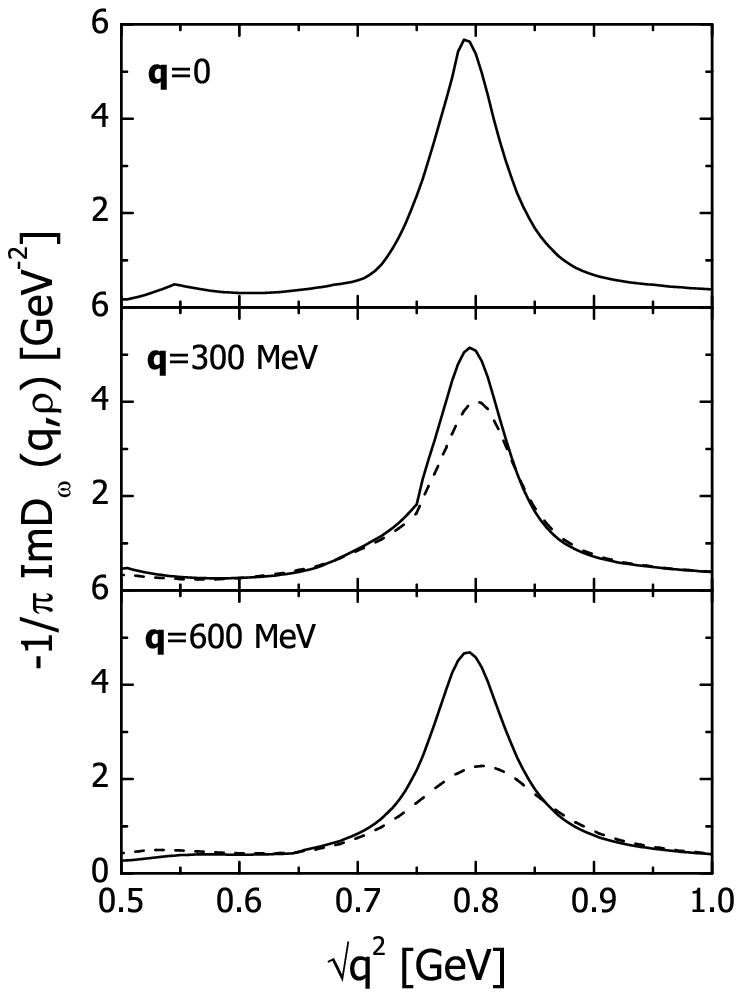}
\caption{The transverse (dashed) and longitudinal (solid) modes of the $\omega$
spectral function at nuclear saturation density $\rho_0=0.16~\mathrm{fm}^{-3}$.
Shown are results for $\omega$ three-momenta $|{\bf q}|=0$,
$|{\bf q}|=300~\mathrm{MeV}$, $|{\bf q}|=600~\mathrm{MeV}$ with respect to nuclear
matter at rest.}
\label{spectral_plab}
\ec\efi
of the $\omega$ spectral function for momenta of
$|{\bf q}|=0~\mathrm{MeV},~300~\mathrm{MeV}$ and 600 MeV.

We observe a significantly different momentum dependence of the two
helicity states:
$\mathcal{A}^T$ is strongly affected at large $\omega$ momenta
whereas $\mathcal{A}^L$ remains almost unchanged.
We note in passing that this resembles qualitatively the spectral functions for the
in-medium $\rho$-meson calculated in \cite{Post:2003hu}.

The same effect is visible in the calculation of the $\omega$ width
evaluated at the actual peak position of the $\omega$ branch. It is
given by the expression
\bea
\Gamma^{L/T}_{\mathrm{peak}}({\bf q})=-\frac{\mathcal{I}m\Pi^{L/T}
\left(q_0=(m_{\mathrm{peak}}^2+{\bf q}^2)^{\frac{1}{2}},{\bf q}\right)}{m_{\mathrm{peak}}}.
\label{widths}
\eea

\bfi\bc
\igr[scale=.8]{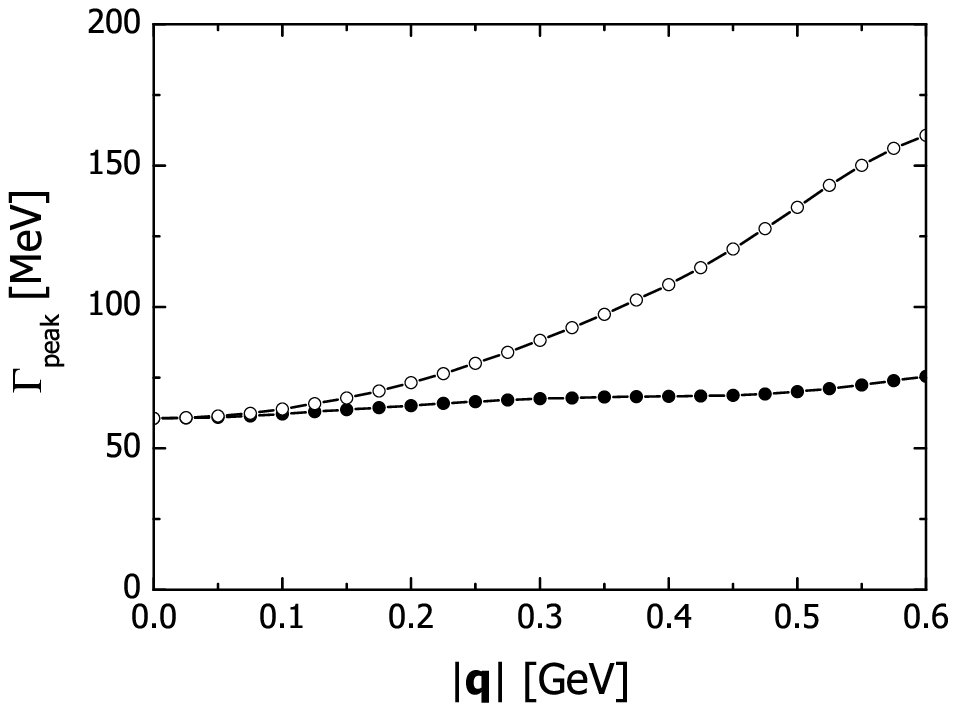}
\caption{The $\omega$ width in nuclear matter at the actual peak position of the
spectral function for different laboratory momenta. Open symbols correspond to
transversely and solid symbols to longitudinally polarized $\omega$ mesons.}
\label{width}
\ec\efi

In Fig.~\ref{width} we show $\Gamma^{L}_{\mathrm{peak}}({\bf q})$ and
$\Gamma^{T}_{\mathrm{peak}}({\bf q})$ as a function of the $\omega$ three-momentum.
For an $\omega$ meson at rest the collisional broadening amounts to
roughly 60 MeV at normal nuclear matter density. For finite three-momentum this width
more or less stays constant for the longitudinal branch whereas it drastically rises
for the transverse modes. From Eqs.~(\ref{self_energ},\ref{widths})
it follows that the widths of the longitudinal and transverse modes,
$\Gamma^{L/T}_{\mathrm{peak}}({\bf q})$, are entirely defined  by the imaginary
parts of the $\omega N$ scattering  amplitudes $\mathcal{I}m\left\{T_{0+\frac{1}{2}}\right\}$ and
$\frac{1}{2}\mathcal{I}m\left\{T_{1-\frac{1}{2}}+ T_{1+\frac{1}{2}}\right\}$, correspondingly (see Appendix \ref{scattering}).
The lower subscript denotes the helicities of the $\omega$-meson and the nucleon.
Note, that at the $\omega N$ threshold  only the $J=\frac{1}{2}^-$
and  $J=\frac{3}{2}^-$ partial waves contribute.
Since the $\omega N$ scattering is dominated by the resonance mechanism
the helicity amplitudes  are governed by the $RN\omega$ coupling
constants extracted in \cite{Shklyar:2004ba}. They are given
in Table~\ref{table01} for completeness.
The different $RN\omega$ coupling constants  correspond
to various helicity combinations of the
$\omega N$ final state, see \cite{Penner:2002ma,Shklyar:2004ba} for details.
With increasing $\omega$ momentum the resonance contributions  become more
important giving  main contributions to the $(1,-\frac{1}{2})$ and $(1,+\frac{1}{2})$
helicity amplitudes. As a result the transverse mode is strongly modified
with increasing $\omega$ momentum.

The peak position of the genuine
$\omega$ branch in both spectral functions moves only slightly to higher
$q^2$ values as can be seen also in Fig.~\ref{peak}. This is due to level
\bfi\bc \igr[scale=.8]{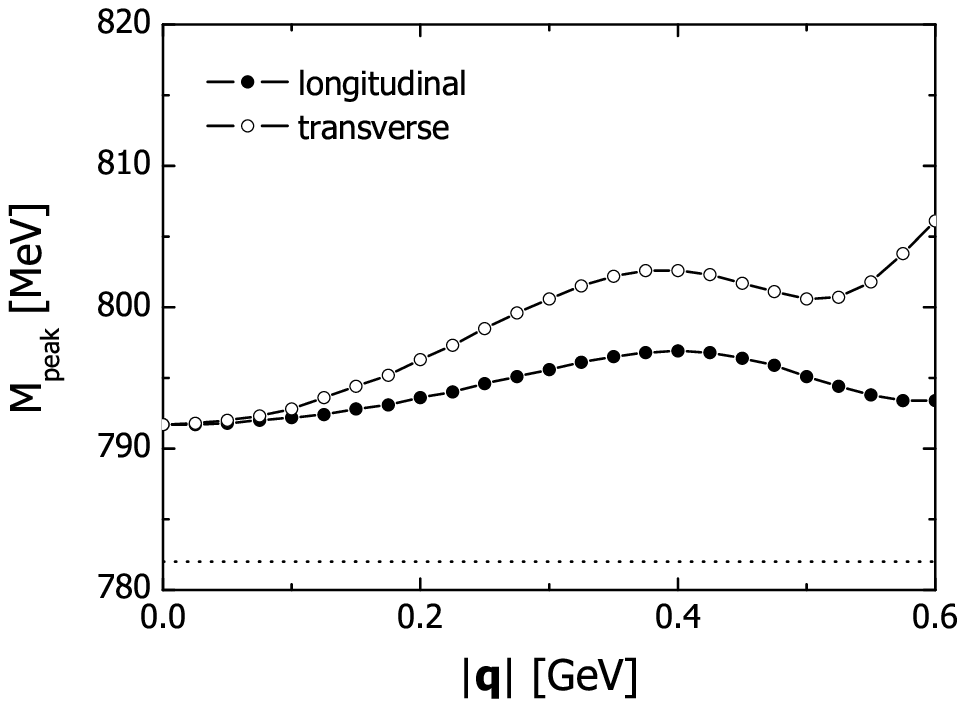} \caption{Peak position  of the
$\omega$ spectral distribution in matter at $\rho_0$. Open symbols
correspond to the transverse and solid symbols to the longitudinal
modes. The dotted line denotes the value of the free $\omega$ peak
mass.} \label{peak} \ec\efi
repulsion as the most important resonance-hole states are
subthreshold with respect to the $N\omega$ channel. Since
the scattering amplitude $T_{\omega N}$ used in our
calculations is a coherent sum of a number of resonance
contributions including coupled-channel effects the separation of
individual resonance contributions is difficult. However, several
conclusion can be drawn.  First, over the full energy range under
consideration the self energy is dominated by resonance hole
excitations, whereas the nucleon Born terms give only marginal
contributions. Although the excitation of
resonance-hole states leads to additional branches in the spectral functions,
no clearly distinguishable peak structures emerge due to the large widths
and the only moderate coupling to the $N\omega$ channel of the individual
resonances. This is also suggested by pion- and photon-induced $\omega$
production data which do not show any prominent resonance structures.
The $\omega N$ amplitude at threshold is dominated by the
$S_{11}(1535)$ resonance and --
through background ($u$-channel) contributions -- by the $D_{15}(1675)$ and $F_{15}(1680)$
states. At zero momentum the $S_{11}(1535)$ resonance generates strength
at low invariant masses. The $u$-channel contributions of the $P_{11}(1710)$ and
$F_{15}(1680)$ resonances that lie only slightly subthreshold to the $N\omega$
channel due to level repulsion push the $\omega$ branch to higher $q^2$ values.

We now compare our results to those obtained from other models. Very close in spirit to our
approach is the work of \cite{Lutz:2001mi} that is also based on a
solution of a coupled-channel Bethe Salpeter equation. This analysis
is restricted to $\omega$ mesons at rest as no $p$-wave resonances have
been incorporated. At least qualitatively the results of \cite{Lutz:2001mi} compare very
well to our findings: Due to resonance-hole excitations an
additional peak in the spectral function was found (however much stronger and
stemming from the unresolved contributions of the $D_{13}(1520)$ and
$S_{11}(1535)$ resonances) whereas the genuine $\omega$ peak is shifted
only slightly upwards in energy.

In \cite{Klingl:1997kf,Klingl:1998zj} the main contribution to the
$\omega$ medium modification comes from a change of the $\rho\pi$
self energy generated by the $\rho N$ and $\pi N$ interactions.
Whereas the on-shell broadening obtained by the authors of Ref.
\cite{Klingl:1998zj} compares very well to our value of roughly 56
MeV, they find an extremely strong attractive mass shift that is not
found in our calculations. In \cite{Klingl:1997kf,Klingl:1998zj} the
real part of the in-medium self energy, that determines the peak
position of the spectral function, is obtained by a dispersion
relation. The magnitude and in particular the energy dependence of
this real part can be attributed to the strong energy variation of
the $\omega N\rightarrow2\pi N$ cross section which in
\cite{Klingl:1998zj} is dominated by the scattering into an
intermediate $\rho N$ state. We note that this $2\pi N$ final
state is not constrained by any data in the calculations
of \cite{Klingl:1998zj}. In contrast, in our approach the $2\pi N$
final state is constrained by the coupled-channel mechanism. In
particular, it has to account for the inelasticity in the
pion-induced reaction channels. This results in a more moderate
energy dependence of the corresponding cross sections and, hence, ---
via dispersion --- in a smaller real part of the scattering amplitude.
We also stress, that the self energy in \cite{Klingl:1998zj} is
obtained from a pure tree level calculation using the heavy-baryon
approximation whose reliability is questionable for the nucleon
lines. Furthermore, we have illustrated in Sect. \ref{sect3} that
the coupling of the $\omega N$ and $\pi N$ channels to nucleon
resonances yields important contributions to the $\omega$ in-medium
spectrum that are absent in the calculation of \cite{Klingl:1998zj}.
In spite of the fact that in our approach the $2\pi N$ final state is constrained by
inelasticity data, one should note that this $2\pi N$ channel as a three-body state is
not treated as rigorously as the two-body 
states ($\pi N$, $\eta N$, $K \Lambda$, ...). This is due to the fact that the inclusion
of a three-particle state in a $K$-matrix approach is much more complicated. Clearly
this leaves some room for further theoretical improvements. 

Recent experiments by the CBELSA/TAPS collaboration
\cite{Trnka:2005ey} have found an in-medium width of the
$\omega$ meson extrapolated to small three-momenta of the $\omega$
of about 50 MeV. This result for the width is in agreement with the
calculations presented here (Fig. \ref{width}). In
\cite{Trnka:2005ey} also a mass shift down to a peak mass of about
720 MeV has been observed. This value has been obtained for the
integrated $\omega$ momentum spectrum from 0 to 500 MeV in this
photoproduction experiment carried out at beam energies from
$0.64~{\rm GeV}$ to $2.53~{\rm GeV}$. The latter result is not in
agreement with our calculations.  This raises the immediate question
for an interpretation of the experimentally observed effect and its
relation to the calculations presented here.

We first note that the observed mass spectrum cannot directly be
compared to the spectral function calculated in this paper since the
experimental results represent a product of the spectral function
with the branching ratio into the $\pi^0\gamma$ channel. The
dominant $\rho \pi$ decay branch increases with mass whereas that
for the $\pi^0 \gamma$ decay is much flatter as a function of the
$\omega$ mass (see Fig. 6 in \cite{Muhlich:2006ps}). Multiplying the
spectral function with the relevant branching ratio thus tends
to shift strength down towards smaller masses. This effect has
not been taken into account in the analysis of the
CBELSA/TAPS experiment. Moreover, the branching ratio is expected
to change in the medium along with the $\rho$ meson properties. We 
also note that the effect observed experimentally obviously 
depends on the background subtraction.

In principle, a similar effect
could also be generated by low-lying resonance-hole excitations that
leave the position of the generic $\omega$ peak almost untouched. Our
calculations presented here contain prominent resonance-hole components 
only at much lower energy (see Fig. \ref{spectral_dens}) associated with the
$N^*(1535)$ nucleon resonance. The absence of higher-lying resonance
excitations is due to the fact that the most recent K-matrix coupled
channel analysis on which the present calculations are based
\cite{Shklyar:2004ba} ascribes most of the cross section in $\gamma
N$ and $\pi N$ reactions at threshold to a combination of various
background terms and small resonance contributions. On the contrary,
the earlier analysis by Penner et al.
\cite{Penner:2002ma,Penner:2002md}, based on older data and not
taking into account the spin-5/2 resonances,  gave a much more
dominant contribution of the $P_{11}$(1710) resonance. Building a
resonance-hole excitation with this state would indeed give a
low-mass component at an energy close to the peak energy of
the $\omega$. This
illustrates the difficulties that arise in predictions of the
$\omega$ self energy in medium due to the still evolving experimental
situation as far as the $\omega$ coupling to nucleon resonances is
concerned. As a consequence, theoretically the real part of the
$\omega$ self energy still is associated with large error bars due to
the inelastic $\omega N$ channels that are hardly constrained
experimentally.

As already indicated in the introduction a discussion of our results in the context of 
in-medium QCD sum rules seems to be expedient. In \cite{Steinmueller:2006id} a sum rule 
analysis of the $\omega$ in-medium spectrum has been done. With this aim a new type of sum 
rule -- the so-called \emph{weighted finite energy sum rule} -- for the study of 
in-medium vector mesons has been established. It has the advantage to exclusively 
relate in-medium hadronic and partonic information instead of mixing in-medium and 
vacuum properties. Moreover, one is essentially free of the problem how to determine a 
reliable Borel window -- an arbitrary mass scale that has to be introduced in order to 
improve the convergence of the involved dispersion integrals. The authors of 
\cite{Steinmueller:2006id} have shown that the sum rules cannot readily determine the 
in-medium spectral shape of the $\omega$ meson. For a given hadronic model of the 
$\omega$ spectral function rather some hadronic parameters can be constrained or 
correlated. Using typical hadronic parametrizations for the $\omega$ in medium spectral 
function a general tendency towards an upwards shift of the $\omega$ mass in the medium 
has been found. This statement, however, strongly depends on the in-medium four-quark 
condensate, a quantity whose density-dependence is far from settled as yet, see also 
Refs.~\cite{Steinmueller:2006id,Thomas:2005dc,Leupold:2005eq} for a discussion of this 
issue. 

A sum rule check of the results presented here is, however, delicate. In the sum rules 
the hadronic information is encoded in the current-current correlation function. 
Usually the correlator in the vector-isoscalar channel is related to the $\omega$ meson 
using strict vector meson dominance (VMD) what means that the interaction of the 
electromagnetic current in this channel with hadrons is saturated by the $\omega$ 
meson. This strict version of VMD is, however, excluded by the sum rules as was shown 
also in \cite{Steinmueller:2006id}. Any other version of VMD needs more microscopic 
input (see for instance \cite{Friman:1997tc}) that is not provided within the 
$K$-matrix approach and, thus, is beyond the scope of the present work. At least 
qualitatively we state that the recent sum rule analysis \cite{Steinmueller:2006id} 
agrees with the results presented here whereas hadronic models that find a strong 
downward shift of the in-medium $\omega$ mass require a rather extreme scenario of 
in-medium changes of the four-quark condensates that can hardly be motivated from our 
present knowledge.

\section{Summary}\label{summary}

Using the low density theorem we have calculated the $\omega$ meson
spectral function at finite nuclear density and zero temperature.
The $\omega N$ forward scattering amplitude is constructed within a unitary
coupled-channel effective Lagrangian model previously applied to the analysis of
pion- and photon-induced reactions on the nucleon. The resulting amplitude is taken
from the updated solution of the coupled-channel problem in the energy region
$\sqrt{s}\leq 2$ GeV. To obtain the spectral function of the $\omega$ meson we have extended
our approach to allow for arbitrary masses and three-momenta of the asymptotic $\omega$ meson
while the intermediate $\omega$ and all other mesons maintain their vacuum
properties. This is in line with the low density approximation.

As a general outcome of our investigations we find that
coupled-channel effects and resonance contributions play an
important role and cannot be neglected when one aims at a reliable
extraction of the $\omega$ in-medium properties. At normal nuclear
density  and zero $\omega$ momentum we find a significant broadening
of about 60 MeV of the $\omega$ spectral function but only a small
upwards shift of the $\omega$ peak relative to the physical $\omega$
mass.

Furthermore, our calculations show that at non-zero momentum the transverse
part of the spectral distribution is affected by the $(1,+\frac{1}{2})$ and $(1,-\frac{1}{2})$ helicity
contributions  from the $\omega N$ channel coming from the resonance part of
the scattering amplitude. This leads to a larger broadening of the transverse
mode of the spectral distribution as compared to the longitudinal one.
The question of an in-medium mass shift of the $\omega$ meson remains to some extent 
an open issue due to the $2\pi N$ state.
In \cite{Klingl:1998zj} it has been found to be responsible for a strong
attractive mass shift. Unfortunately up to the present three-body states
cannot be treated in a rigorous way in the $K$-matrix approach.
Nonetheless we would like to stress again that also such exotic
reactions as $\omega N \to 2\pi N$ should be constrained by data as much as
possible - if it turns out that such a channel is important for
in-medium modifications.
The inclusion of this three-body final
state into coupled-channel $K$-matrix calculations is a highly non-trivial task that
will be subject to future investigations.

This work has been supported by DFG and BMBF. We thank M.~Lutz and
W.~Weise for discussions on the subject.


\appendix
\section{The $\omega$ meson in vacuum}\label{vacuum}

\begin{table}[hbt!]
\bc
\begin{tabular}{c|c|c}
\hline\hline
channel & ratio [\%] & width [MeV] \\
\hline\hline
$\pi^+\pi^-\pi^0$ & 89.6 & 7.56 \\
\hline
$\pi^0\gamma$ & 8.7 & 0.73 \\
\hline
$\pi\pi$ & 1.7 & 0.14 \\
\hline\hline
\end{tabular}
\caption{Decay channels, branching ratios and partial widths of the $\omega$ meson in vacuum \cite{Eidelman:2004wy}. The value for the $\omega\rightarrow 3\pi$ width is chosen somewhat larger as in \cite{Eidelman:2004wy} (but still within the error bars) in order to saturate the total width by the three dominating decay channels.}
\label{table02}
\ec
\end{table}

The hadronic $\omega$ vacuum self energy is given as a sum of the contributions coming from the
coupling of the $\omega$ to the channels $3\pi$, $\pi^0\gamma$ and $2\pi$, adding up to
give a total vacuum decay width of about 8.44 MeV. The partial widths are given in
Table~\ref{table02}. We assume that the
$\omega\rightarrow 3\pi$ decay proceeds via an intermediate $\rho$ meson,
i.e.~$\omega\rightarrow\rho\pi\rightarrow 3\pi$. For the corresponding decay width we find
\bea
\Gamma_{\omega\rightarrow 3\pi}(q^2)=\int\limits_{4m_{\pi}^2}^{(\sqrt{q^2}-m_{\pi})^2}ds
\Gamma_{\omega\rightarrow\rho\pi}(q^2,s)\mathcal{A}_{\rho}(s)
\frac{\Gamma_{\rho\rightarrow\pi\pi}(s)}{\Gamma_{\rho}^{\mathrm{tot}}(s)}
\eea
with the $\rho$ vacuum spectral function $\mathcal{A}_{\rho}$. From the Lagrangians used
in our model \cite{Penner:PhD} we obtain
\bea
\Gamma_{\omega\rightarrow\rho\pi}(q^2,s)=\frac{3g^2}{4\pi}\frac{p(q^2,s)^3}{m_{\pi}^2},
\eea
where $p(q^2,s)$ is the center of mass momentum of a $\rho$ meson of mass $\sqrt{s}$ from
the decay of an $\omega$ of mass $\sqrt{q^2}$ into the $\rho\pi$ system. The two-pion
decay width of the $\rho$ meson is given by
\bea
\Gamma_{\rho\rightarrow\pi\pi}(s)=\Gamma_0\left(\frac{m_{\rho}}{\sqrt{s}}\right)^2
\left(\frac{s-4m_{\pi}^2}{m_{\rho}^2-4m_{\pi}^2}\right)^{\frac{3}{2}}\Theta(s-4m_{\pi}^2)
\eea
with the on-shell decay width $\Gamma_0=149.2$ MeV.

The $\omega\rightarrow\pi\pi$ decay
width is given by the very same expression but with the $\rho$ on-shell width replaced
by the corresponding decay width of the $\omega$. From \cite{Klingl:1996by} we adopt the
width for the semi-hadronic
decay $\omega\rightarrow\pi^0\gamma$:
\bea
\Gamma_{\omega\rightarrow\pi^0\gamma}(q^2)=\frac{9}{24\pi}
\left(\frac{d}{f_{\pi}}\right)^2\left(\frac{q^2-m_{\pi}^2}{\sqrt{q^2}}\right)^3
\Theta(q^2-m_{\pi}^2)
\eea
with $d\simeq 0.1$ and the pion decay constant $f_{\pi}=92.4$ MeV. Neglecting the real
part of the $\omega$ vacuum self energy, i.~e. using the physical $\omega$ mass in the
vacuum spectral function, we obtain the $\omega$ vacuum self energy:
\bea
\Pi(q^2)=-i\sqrt{q^2}\left(\Gamma_{\omega\rightarrow 3\pi}(q^2)
+\Gamma_{\omega\rightarrow\pi^0\gamma}(q^2)+\Gamma_{\omega\rightarrow 2\pi}(q^2)\right) \,.
\eea


\section{Lagrangians}\label{lagrangians}

As the coupling of the $N\omega$ channel to nucleon resonances is of special importance
for the evaluation of the $\omega$ in-medium self energy, we give in the following the
corresponding Lagrangians entering the $K$-matrix interaction potential:
\begin{eqnarray}
\mathcal{L}_{\frac{1}{2}N\omega} &=& -\bar u_R\left\{1\atop -i\gamma_5 \right\}
\left(g_1\gamma_{\mu}-\frac{g_2}{2m_N}\sigma_{\mu\nu}\partial_{\omega}^{\nu}\right)
u_N \omega^{\mu} + h.c.~, \\ \label{L32}
\mathcal{L}_{\frac{3}{2}N\omega} &=& -\bar u_R^{\mu}\left\{ i\gamma_5\atop 1\right\}
\left(\frac{g_1}{2m_N}\gamma^{\alpha}+i\frac{g_2}{4m_N^2}\partial_N^{\alpha}
+i\frac{g_3}{4m_N^2}\partial_{\omega}^{\alpha}\right)
\left(\partial^{\omega}_{\alpha}g_{\mu\nu}-\partial^{\omega}_{\mu}g_{\alpha\nu}\right)
u_N \omega^{\nu}\nonumber\\ && + h.c.~,  \\
\mathcal{L}_{\frac{5}{2}N\omega} &=& \bar u_R^{\mu\lambda}
\left\{1\atop i\gamma_5 \right\}\left(\frac{g_1}{4m_N^2}\gamma^{\xi}
+i\frac{g_2}{8m_N^3}\partial_N^{\xi}+i\frac{g_3}{8m_N^3}\partial_{\omega}^{\xi}\right)
\left(\partial^{\omega}_{\xi}g_{\mu\nu}-\partial^{\omega}_{\mu}g_{\xi\nu}\right)
u_N\partial^{\omega}_{\lambda}\omega^{\nu} \nonumber\\ && + h.c.~,
\end{eqnarray}
where $h.c.$ denotes the hermitian conjugate.
In all three cases the upper operator holds for positive and the lower one for negative parity resonances. In the 
spin-$\frac{3}{2}$ case the vertices are contracted with an off-shell projector that, for simplicity, is not displayed in Eq.~(\ref{L32}), see e.~g. \cite{Penner:2002ma}. One should keep in mind, that the $\omega N$ forward scattering amplitude is not obtained by just summing the individual tree-level contributions of the included resonances. In fact, by means of the Bethe-Salpeter equation rescattering effects are taken into account that turn out to play an important role already in the description of photon and pion induced $\omega$ production data on the nucleon.


\section{Scattering length}\label{scattering}

The definition of the scattering length in the present paper differs from
the definitions used in \cite{Shklyar:2004ba} and \cite{Lutz:2001mi}. There
 a decomposition of the $\omega N$ helicity  amplitudes with respect to
the total angular momentum of the $\omega N$ system has been performed.
In the present paper, however, we define the scattering length as done
in \cite{Klingl:1998zj}, what is consistent with the evaluation of the
self energy as given by Eqs.~(\ref{project}) and (\ref{trho}):
\bea
a_{\omega N}=\frac{m_N}{4\pi(m_N+m_{\omega})}T_{\omega N}(q_0=m_{\omega}),
\eea
where $T_{\omega N}$ is the spin- and helicity-averaged $\omega N$ forward
scattering amplitude at threshold:

\bea
T_{\omega N}(m_{\omega})=\frac{1}{2}\left(T_{+1+\frac{1}{2}}
(m_{\omega})+T_{+1-\frac{1}{2}}(m_{\omega})\right)=
T_{0+\frac{1}{2}}(m_{\omega}).
\label{om_average}
\eea
The lower  indices stand for  the $\omega$ and nucleon helicities.
The amplitudes in the right-hand side of the Eq.(\ref{om_average})
 are obtained from the partial wave decomposition \cite{Penner:2002ma}
\bea
T_{\lambda}(m_\omega) &=&\frac{4\pi(m_N+m_\omega)}{p\, m_N}\sum_{J}
\left(J+\frac{1}{2}\right)d_{\lambda\lambda'}^J
(0)\left( T_{\lambda'\lambda}^{J+}(m_N+m_\omega) +
T_{\lambda'\lambda}^{J-}(m_N+m_\omega)\right)\nonumber\\
&=&\frac{4\pi(m_N+m_\omega)}{p\, m_N}
\left( T_{\lambda\lambda}^{\frac{1}{2}-}(m_N+m_\omega)
+ 2  T_{\lambda\lambda}^{\frac{3}{2}-}(m_N+m_\omega) \right),
\eea
where $p$ is the c.m. three-momentum and $\lambda = \lambda_\omega +\lambda_N$.
Note, that only the $J^P=\frac{1}{2}^-$ and  $J^P=\frac{3}{2}^-$ partial waves
contribute close to the $\omega N$ threshold.

With the definition (\ref{om_average}), the classical interpretation of the scattering
length similar as for spinless particles holds:
\bea
\sigma(\sqrt{s}=m_N+m_{\omega})=\sigma^{\frac{1}{2}}+\sigma^{\frac{3}{2}}=
4\pi\left(3|a_{\omega N}^{\frac{1}{2}^-}|^2+\frac{3}{2}|a_{\omega N}^{\frac{3}{2}^-}|^2\right)
\eea
where $\sigma$ is the usual spin- and helicity-averaged
total $\omega N$ elastic cross section at threshold. With this definition
the following formula for the on-shell mass shift applies:
\bea
\Delta m=-\frac{2\pi\rho_N}{m_{\omega}}\left(1+\frac{m_{\omega}}{m_N}\right)
\mathcal{R}e~a_{\omega N}
\eea
yielding a value of roughly 15 MeV. Note, however, that the shift of
the $\omega$ peak in the spectral function is somewhat smaller since
the real part of the self energy is reduced for $q^2$ values slightly
above the $\omega$ pole mass, see Fig.~\ref{self}.

\bibliography{biblio_omega}
\end{document}